\documentclass{article}
\usepackage{amsfonts}
\usepackage{amssymb}
\usepackage{amsmath}
\usepackage{fancyhdr}
\usepackage[symbol]{footmisc}

\newtheorem{theorem}{Theorem}

\newtheorem{proposition}[theorem]{Proposition}

\newenvironment{proof}[1][Proof]{\noindent\textbf{#1.} }{\ \rule{0.5em}{0.5em}}
\input{tcilatex}
\begin{document}

\begin{center}
{\large \textbf{Lagrangian Description, Symplectisation }}

{\large \textbf{and Eulerian Dynamics of Incompressible Fluids\\[2mm]
}}H. G\"{u}mral

Department of Mathematics, Yeditepe University

34755 Ata\c{s}ehir, Istanbul, Turkey

\ hgumral@yeditepe.edu.tr \\[0.5cm]
\end{center}

\textbf{Abstract:}Eulerian dynamical equations on three dimensional domain
are used to construct a formal symplectic structure on time-extended space.
Symmetries, invariants and conservation laws are related to this geometric
structure. Symplectic structure incorporates dynamics of helicities as
identities. Generator of the infinitesimal dilation for symplectic two-form
can be interpreted as a current vector for helicity. Symplectic dilation
implies existence of contact hypersurfaces. In particular, these include
contact structures on the space of streamlines and on the Bernoulli surfaces.

\section{Introduction}

In this work, we shall be concerned with the relations between Lagrangian
description and Eulerian equations of incompressible fluid and, exploit the
Eulerian evolution equations of some hydrodynamic systems to obtain
geometric structures relevant to a qualitative study of the Lagrangian
description of motion in three dimensions.

The Eulerian equations can be used to define an exact symplectic structure
on time-extended space of trajectories of dynamical field. For
incompressible fluids, the suspended velocity field as well as the vorticity
field with a multiplicative factor can be realized as Hamiltonian vector
fields with respect to this structure with the Hamiltonian functions given
by the Bernoulli function and time, respectively.

One can then relate symmetries, invariant differential forms and
conservation laws within the framework of this geometric structure
intrinsically contained in the Eulerian equations. In particular, the
symplectic structure turns out to have a prominant role in relating the
vorticity field to the helicity invariant.

The helicity conservation law can be interpreted as the
symplectic-divergence of a vector field associated with the symplectic form,
namely, the symplectic dilation or the Liouville vector field. Although, it
is not a Hamiltonian vector field, Liouville vector field generates, via Lie
derivative, Hamiltonian vector fields from Hamiltanian vector fields. This
enables us to identify three dimensional domains in which the dynamical
field admits infinitely many symmetries of kinematical type.

A hypersurface in a symplectic manifold admits a contact one-form if and
only if there exists a symplectic dilation which is defined on its
neighborhood and is transversal to the hypersurface. It turns out that a
current vector field governing the dynamics of the helicity density
generates the dilation for the symplectic structure.

For fluid dynamical content of this work we shall refer to Refs. \cite{CM}-%
\cite{MEY} and the necessary mathematical background can be found in Refs. 
\cite{JMS}-\cite{MR}.

\section{Eulerian Dynamics}

\textbf{1.} \textbf{Euler Equations:} We shall begin with the Euler
equations of ideal incompressible fluids 
\begin{equation}
{\frac{\partial \mathbf{v}}{\partial t}}+(\mathbf{v}\cdot \nabla )\mathbf{v}%
=-\nabla p  \label{euler}
\end{equation}%
for the divergence-free velocity field $\mathbf{v}$ tangent to the boundary
of a connected region $M\subset 
\mathbb{R}
^{3}$ with coordinates $\mathbf{x}$ and the pressure function $p$. The
identity ($\mathbf{v}\cdot \nabla )\mathbf{v}=\nabla |\mathbf{v}|^{2}/2-%
\mathbf{v}\times (\nabla \times \mathbf{v})$ can be used to bring the Euler
equation (\ref{euler}) into the Bernoulli's form 
\begin{equation}
{\frac{\partial \mathbf{v}}{\partial t}}-\mathbf{v}\times (\nabla \times 
\mathbf{v})=-\nabla \alpha  \label{reuler}
\end{equation}%
where the function $\alpha \equiv p+v^{2}/2$ is the Bernoulli function \cite%
{BAK}, also called the total or stagnation pressure \cite{MEY}. In terms of
the divergence-free vorticity field $\mathbf{w}\equiv \nabla \times \mathbf{v%
}$ Eq.(\ref{reuler}) gives 
\begin{equation}
{\frac{\partial \mathbf{w}}{\partial t}}-\nabla \times (\mathbf{v}\times 
\mathbf{w})=0\;.  \label{weq}
\end{equation}%
It follows from the identity $\nabla \times (\mathbf{v}\times \mathbf{w})=(%
\mathbf{w}\cdot \nabla )\mathbf{v}-(\mathbf{v}\cdot \nabla )\mathbf{w}%
+(\nabla \cdot \mathbf{w})\mathbf{v}-(\nabla \cdot \mathbf{v})\mathbf{w}$
together with $\nabla \cdot \mathbf{v}=\nabla \cdot \mathbf{w}=0$ that Eq.(%
\ref{weq}) is equivalent to 
\begin{equation}
{\frac{\partial \mathbf{w}}{\partial t}}+[\mathbf{v},\mathbf{w}]=0\;,\;\;\;[%
\mathbf{v},\mathbf{w}]\equiv (\mathbf{v}\cdot \nabla )\mathbf{w}-(\mathbf{w}%
\cdot \nabla )\mathbf{v}  \label{symm}
\end{equation}%
which means that $\mathbf{w}$ is an infinitesimal time-dependent symmetry of
the velocity field $\mathbf{v}$. Solutions $\mathbf{x=x(}t\mathbf{)}$ of the
ordinary differential equations $d\mathbf{x}/dt=\mathbf{v}(\mathbf{x},t)$
are tarjectories of the velocity field. The time-dependent transformations
generated by $\mathbf{w}$ on $M$ leaves these trajectories invariant. A
time-dependent conserved function for the velocity field can be found again
from the Euler equation. We recall that an energy consequence of the Euler
equation follows by taking dot product of its Bernoulli form with the
velocity field. The result is known as the Bernoulli equation \cite{FRI},%
\cite{MEY} 
\begin{equation}
{\frac{\partial }{\partial t}}({\frac{1}{2}}v^{2})+\mathbf{v}\cdot \nabla
\alpha =0
\end{equation}%
which implies that if the pressure $p$ does not depend explicitly on time,
that is, if $p=p(\mathbf{x})$, \ then the Bernoulli function $\alpha $ is a
time-dependent conserved function 
\begin{equation}
{\frac{\partial \alpha }{\partial t}}+\mathbf{v}\cdot \nabla \alpha
=0\;,\;\;\;\alpha (t,\mathbf{x})=p(\mathbf{x})+{\frac{1}{2}}v^{2}(t,\mathbf{x%
})
\end{equation}%
along the trajectories of the velocity field.

\bigskip \textbf{2.} \textbf{Navier-Stokes Equations:} The Navier-Stokes
equations for a viscous incompressible fluid in a bounded domain $M\subset 
\mathbb{R}
^{3}$ is%
\begin{equation}
{\frac{\partial \mathbf{v}}{\partial t}}+(\mathbf{v}\cdot \nabla )\mathbf{v}%
=-\nabla p+\nu \nabla ^{2}\mathbf{v}  \label{nav}
\end{equation}%
where $\nu $ is the kinematic viscosity \cite{FRI}. Eq.(\ref{nav}) results
in the equation 
\begin{equation}
{\frac{\partial \mathbf{w}}{\partial t}}-\nabla \times (\mathbf{v}\times 
\mathbf{w})=\nu \nabla ^{2}\mathbf{w}
\end{equation}%
for the divergence-free vorticity field. Corresponding to the Bernoulli
equation of Euler flow, we have, for the viscous case%
\begin{equation}
{\frac{\partial }{\partial t}}({\frac{1}{2}}v^{2})+\mathbf{v}\cdot \nabla
\alpha =\nu \mathbf{v\cdot }\nabla ^{2}\mathbf{v}
\end{equation}%
from which it follows that%
\begin{equation}
{\frac{\partial \alpha }{\partial t}}+\mathbf{v}\cdot \nabla \alpha ={\frac{%
\partial p}{\partial t}}+\nu \mathbf{v\cdot }\nabla ^{2}\mathbf{v}\text{ .}
\end{equation}%
Thus, if the time dependence of pressure is given by%
\begin{equation}
{\frac{\partial p}{\partial t}}+\nu \mathbf{v\cdot }\nabla ^{2}\mathbf{v=}0
\label{pressure}
\end{equation}%
then the Bernoulli function becomes a conserved quantity of the viscous
Lagrangian flow as well.

\textbf{3.} \textbf{Generalities on Eulerian Equations:} More generally, we
can consider evolution equation 
\begin{equation}
{\frac{\partial \mathbf{v}}{\partial t}}+(\mathbf{v}\cdot \nabla )\mathbf{v}=%
\mathbf{F}  \label{fsymp}
\end{equation}%
for an arbitrary force field $\mathbf{F}$. To account magnetohydrodynamic
systems, we can replace the vorticity by a divergence-free frozen-field $%
\mathbf{B}$ satisfying the condition in Eq.(\ref{symm}). In this case, if
the force field $\mathbf{F}$ satisfies the condition $\mathbf{B}\cdot (%
\mathbf{F}+v^{2}/2)=0$, then the function $\mathbf{v}\cdot \mathbf{B}$ turns
out to be an invariant of the Lagrangian flow \cite{hg97}.

With this remark, following discussions can be extended to the Eulerian
equations such as the equations describing the Boussinesq approximation to
inhomogeneous Euler equations, equations of barotropic fluids, equations of
non-relativistic superconductivity, equations of ideal magnetohydrodynamics
and dynamo theory (see \cite{hghel} and the references therein).

\section{Geometry}

\textbf{4. Symplectic Two-form:} We will obtain from the Eulerian equations
a non-degenerate closed two-form $\Omega _{\nu }$ on $\mathcal{%
\mathbb{R}
\times }M$, that is, a symplectic structure \cite{JMS},\cite{AM},\cite{MR},%
\cite{ARN},\cite{LM}.

\begin{proposition}
Let $\phi $ be an a priori unspecified function on $\mathcal{%
\mathbb{R}
\times }M$. Then, the two-form 
\begin{equation}
\Omega _{\nu }=\mathbf{w}\cdot (d\mathbf{x}\wedge d\mathbf{x})-(\mathbf{v}%
\times \mathbf{w}+\nabla \phi +\nu \nabla ^{2}\mathbf{v})\cdot d\mathbf{x}%
\wedge dt  \label{nsym}
\end{equation}%
on $\mathcal{%
\mathbb{R}
\times }M$ is symplectic on the space of solutions of the Navier-Stokes
equations provided $\mathbf{w}\cdot (\nu \nabla ^{2}\mathbf{v}+\nabla \phi
)\neq 0$.

\begin{proof}
The three-form 
\begin{equation}
d\Omega _{\nu }=(\nabla \cdot \mathbf{w})\text{ }(d\mathbf{x}\wedge d\mathbf{%
x}\wedge d\mathbf{x})+[\mathbf{w}_{,t}-\nabla \times (\mathbf{v}\times 
\mathbf{w-}\nu \nabla \times \mathbf{w})]\cdot d\mathbf{x}\wedge d\mathbf{x}%
\wedge dt
\end{equation}%
vanishes for divergence-free vector field $\mathbf{w}$ satisfying the
Navies-Stokes equations in the vorticity form (\ref{weq}). So, $\Omega _{\nu
}$ is closed. For $\nu =0$ this reduces to the two-form 
\begin{equation}
\Omega _{e}=\mathbf{w}\cdot (d\mathbf{x}\wedge d\mathbf{x})-(\mathbf{v}%
\times \mathbf{w}+\nabla \phi )\cdot d\mathbf{x}\wedge dt\;,  \label{symp2}
\end{equation}%
which is closed by the Euler equations in rotational form. For
non-degeneracy, we compute%
\begin{equation}
{\frac{1}{2}}\Omega _{\nu }\wedge \Omega _{\nu }=(\nu \mathcal{H}_{w}-%
\mathbf{w}\cdot \nabla \phi )\;d\mathbf{x}\wedge d\mathbf{x}\wedge d\mathbf{x%
}\wedge dt\neq 0  \label{nvol}
\end{equation}%
where the scalar function 
\begin{equation}
\mathcal{H}_{w}\equiv \mathbf{w}\cdot \nabla \times \mathbf{w}=-\mathbf{w}%
\cdot \nabla ^{2}\mathbf{v}
\end{equation}%
is known as the vortical helicity density \cite{FRI}. For viscous flows $%
\nabla ^{2}\mathbf{v}\neq 0$ and, for a realistic fluid we have $\mathbf{w}%
\neq 0$ \cite{saf81}. Hence $\mathcal{H}_{w}\neq 0$. This makes $\Omega
_{\nu }$ non-degenerate. For the Euler equations, Eq. (\ref{nvol}) reduces to%
\begin{equation}
{\frac{1}{2}}\Omega _{e}\wedge \Omega _{e}=-(\mathbf{w}\cdot \nabla \phi )\;d%
\mathbf{x}\wedge d\mathbf{x}\wedge d\mathbf{x}\wedge dt\neq 0  \label{vol}
\end{equation}%
which, for non-degeneracy, requires $\nabla \phi $ to be non-zero.
\end{proof}
\end{proposition}

The non-zero four-form in Eq.(\ref{vol}) is the symplectic or the Liouville
volume element on $\mathcal{%
\mathbb{R}
\times }M$.

\bigskip \textbf{5. Exactness:} For a one-form $\theta =\psi dt+\mathbf{v}%
\cdot d\mathbf{x}$ we compute%
\begin{equation*}
d\theta =\mathbf{w}\cdot (d\mathbf{x}\wedge d\mathbf{x})-(\mathbf{v}\times 
\mathbf{w}+\nabla (\psi +\alpha )-\nu \nabla \times \mathbf{w})\cdot d%
\mathbf{x}\wedge dt
\end{equation*}%
where we solved the time derivative of the velocity field from the
Navier-Stokes equations. The right-hand-side is the same as the symplectic
form provided we have $\psi +\alpha =-\phi $. Thus, the parametric family of
symplectic two-forms is exact 
\begin{equation*}
\theta =-(\phi +\alpha )dt+\mathbf{v}\cdot d\mathbf{x,}\ \ \ \Omega _{\nu
}=d\theta \text{ \ }\ mod\ Eq.(\ref{nav})
\end{equation*}%
and this includes $\Omega _{e}=d\theta $ \ $\ mod\ Eq.(\ref{euler})$ for the
Euler equations.

\textbf{6. Hamiltonian Vector Fields:} The non-degeneracy of $\Omega _{\nu }$
means that given a one-form $\beta \equiv \beta _{a}dx^{a}$ on $\mathcal{%
\mathbb{R}
\times }M$ with the local coordinates $(x^{a})=(x^{0}=t,\mathbf{x})$ the
equations 
\begin{equation}
i(X)(\Omega _{\nu })=\beta \;,\;\;\;(\Omega _{\nu })_{ab}X^{a}=\beta _{b}
\label{flat}
\end{equation}%
has a unique solution for the vector field $X=X^{a}\partial
_{a}=X^{0}\partial _{t}+\mathbf{X}\cdot \nabla $ and vice versa. Here, $%
i(X)(\cdot )$ denotes the interior product or the contraction with the
vector field $X$ \cite{AMR}, $(\Omega _{\nu })_{ab}$ are the components of
the skew-symmetric matrix of the symplectic two-form $\Omega _{\nu }$ in the
given coordinates and we employ the summation over repeated indices.

For an arbitrary smooth function $f$ on $\mathcal{%
\mathbb{R}
\times }M$ the Hamiltonian vector field $X_{f}$ defined by the symplectic
two-form $\Omega _{\nu }$ is 
\begin{equation}
X_{f}={\frac{1}{\nu \mathcal{H}_{w}-\mathbf{w}\cdot \nabla \phi }}[{\frac{df%
}{dt}}\mathbf{w}\cdot \nabla -(\mathbf{w}\cdot \nabla f)({\frac{\partial }{%
\partial t}}+\mathbf{v}\cdot \nabla )+(\nabla \phi -\nu \nabla \times 
\mathbf{w})\times \nabla f\cdot \nabla ]
\end{equation}%
and this satisfies the Hamilton's equations $i(X_{f})(\Omega _{\nu })=df$.
Here, $d/dt=\partial /\partial t+\mathbf{v}\cdot \nabla $ is the convective
(or material) derivative. From the skew-symmetry of the matrix $(\Omega
_{\nu })_{ab}$ we have the conservation law 
\begin{equation}
0\equiv i(X_{f})i(X_{f})(\Omega _{\nu })=(\Omega _{\nu
})_{ab}X^{a}X^{b}=X^{a}f_{,a}=X^{0}f_{,t}+\mathbf{X}\cdot \nabla f=0
\label{cons}
\end{equation}%
for the Hamiltonian function. In particular, for the function $f=t$ we
obtain the Hamiltonian vector field

\begin{equation*}
i(X_{t})(\Omega _{\nu })=dt\text{ \ \ \ \ \ \ \ \ }X_{t}=(\nu \mathcal{H}%
_{w}-\mathbf{w}\cdot \nabla \phi )^{-1}\mathbf{w}\cdot \nabla
\end{equation*}%
which is the vorticity field normalized with the Liouville volume.

\textbf{7. Poisson Bracket:} The symplectic structure $\Omega _{\nu }$ on $%
\mathcal{%
\mathbb{R}
\times }M$ induces a Lie algebraic structure on the space of smooth
functions on $\mathcal{%
\mathbb{R}
\times }M$ with the Poisson bracket 
\begin{eqnarray}
\{f,g\}_{\nu } &\equiv &{\frac{\partial f}{\partial x^{a}}}(\Omega _{\nu
}^{-1})^{ab}{\frac{\partial g}{\partial x^{b}}}\;=-X_{f}(g)\;=\;\Omega _{\nu
}(X_{f},X_{g}) \\
&=&{\frac{1}{\nu \mathcal{H}_{w}-\mathbf{w}\cdot \nabla \phi }}[{\frac{dg}{dt%
}}\mathbf{w}\cdot \nabla f-{\frac{df}{dt}}\mathbf{w}\cdot \nabla g-(\nabla
f\times \nabla g)\cdot \nabla \phi \\
&&\text{ \ \ \ \ \ \ \ \ \ \ \ \ \ \ \ \ \ \ \ \ \ \ \ \ \ \ \ \ }+\nu
(\nabla f\times \nabla g)\cdot (\nabla \times \mathbf{w)}]  \notag
\end{eqnarray}%
where $(\Omega _{\nu }^{-1})^{ab}$ are components of the inverse of the
matrix of symplectic two-form. Skew-symmetry of $(\Omega _{\nu })_{ab}$
implies that $\{,\}_{\nu }$ is skew-symmetric and the fact that $\Omega
_{\nu }$ is closed corresponds to the Jacobi identity%
\begin{equation*}
\{\{f,g\}_{\nu },h\}_{\nu }+\{\{h,f\}_{\nu },g\}_{\nu }+\{\{g,h\}_{\nu
},f\}_{\nu }=0
\end{equation*}%
for arbitrary functions $f,g,h$ on $\mathcal{%
\mathbb{R}
\times }M$. The Lie algebra isomorphism 
\begin{equation}
\lbrack X_{f},X_{g}]=-X_{\{f,g\}}
\end{equation}%
between the algebra of Hamiltonian vector fields and the Poisson bracket
algebra of functions is induced by the symplectic structure $\Omega _{\nu }$ 
\cite{MR},\cite{LM}.

\textbf{8. Helicity Conservation:} A conservation law for Eulerian equations
is a divergence expression of the form $\partial T/\partial t+\nabla \cdot 
\mathbf{P}=0$ with $T$ being conserved density and $\mathbf{P}$ the
corresponding flux. We call the four-vector $(T,\mathbf{P})$ on $\mathcal{%
\mathbb{R}
\times }M$ to be the current associated with the conserved quantity $T$. We
will consider some evolution equations which may be related to Eulerian
conservation laws under certain conditions. These are derived from the
formal symplectic structure as differential identities and will include
evolution equations for helicity and potential vorticity.

Since $\Omega _{\nu }$ is closed and $d\theta =\Omega _{\nu }$, these
differential forms satisfy the relation 
\begin{equation}
d(\theta \wedge \Omega _{\nu })-\Omega _{\nu }\wedge \Omega _{\nu }\equiv 0
\label{tom}
\end{equation}%
identically. Here, we compute the three-form 
\begin{eqnarray}
\theta \wedge \Omega _{\nu } &=&\mathcal{H}\,d\mathbf{x}\wedge d\mathbf{x}%
\wedge d\mathbf{x}+ \\
&&((v^{2}-\phi -\alpha )\mathbf{w}-\mathcal{H}\mathbf{v}-\mathbf{v}\times
(\nabla \phi -\nu \nabla \times \mathbf{w)})\cdot d\mathbf{x}\wedge d\mathbf{%
x}\wedge dt  \notag
\end{eqnarray}%
where the scalar component, namely, the coefficient of the term $d\mathbf{x}%
\wedge d\mathbf{x}\wedge d\mathbf{x}$ is the helicity density 
\begin{equation}
\mathcal{H}\equiv \mathbf{v}\cdot \nabla \times \mathbf{v}=\mathbf{v}\cdot 
\mathbf{w}\;.  \label{pot}
\end{equation}

\begin{proposition}
$\mathcal{H}$ is an Eulerian conserved quantity for the Euler equation ($\nu
=0$) and, for the Navier-Stokes equation if $\mathcal{H}_{w}=0$.

\begin{proof}
We recall Eq.(\ref{tom}) which results in the divergence expression%
\begin{equation}
{\frac{\partial \mathcal{H}}{\partial t}}+\nabla \cdot (\mathcal{H}\mathbf{%
v+(}p-{\frac{1}{2}}v^{2})\mathbf{w})=\nu (\mathbf{v\cdot }\nabla ^{2}\mathbf{%
\mathbf{w-}}\mathcal{H}_{w})  \label{helicity}
\end{equation}%
for the evolution of helicity. Note that, above equation is independent of
the function $\phi $ introduced artificially in the definition of $\Omega
_{\nu }$. Eq.(\ref{helicity}) can also be written in the form%
\begin{equation}
{\frac{\partial \mathcal{H}}{\partial t}}+\nabla \cdot (\mathcal{H}\mathbf{%
v+(}p-{\frac{1}{2}}v^{2})\mathbf{w}-{\nu }\mathbf{v}\times \mathbf{(}\nabla 
\mathbf{\times \mathbf{w})})=-2\nu \mathcal{H}_{w}  \label{helcon}
\end{equation}%
from which one can conclude the conservation of helicity even for viscous
fluids provided the vortical helicity vanishes. In the case of the Euler
equations, we have $\nu =0$ and Eq.(\ref{helcon}) implies%
\begin{equation}
{\frac{\partial \mathcal{H}}{\partial t}}+\nabla \cdot (\mathcal{H}\mathbf{%
v+(}p\mathbf{-}{\frac{1}{2}}v^{2})\mathbf{w})=0  \label{orhe}
\end{equation}%
for the conservation law of the helicity density $\mathcal{H}$ without
further assumption.
\end{proof}
\end{proposition}

\textbf{9. Further Conservation Laws: }Although, the conservation of
Bernoulli's function requires some additional conditions to hold, we can
obtain, for incompressible fluids in general, an Eulerian conservation law
involving Bernoulli's function and the vorticity by considering the
three-form $d\alpha \wedge \Omega _{\nu }$.

\begin{proposition}
The potential vorticities $\mathbf{w}\cdot \nabla \alpha $ and $\mathbf{w}%
\cdot \nabla \phi $ are Eulerian conservation laws for incompressible
viscous fluids.

\begin{proof}
The identity $d(d\alpha \wedge \Omega _{\nu })\equiv 0$ gives the evolution
equation%
\begin{equation*}
\frac{\partial (\mathbf{w}\cdot \nabla \alpha )}{\partial t}+\nabla \cdot %
\left[ (\mathbf{w}\cdot \nabla \alpha )\mathbf{v}+\nu \mathbf{v}\cdot \nabla
\times \mathbf{w+}\nu \nabla \alpha \times (\nabla \times \mathbf{w)}\right]
=0
\end{equation*}%
for the conservation of the potential vorticity $\mathbf{w}\cdot \nabla
\alpha $. Similarly, for $\mathbf{w}\cdot \nabla \phi $ the conservation law
reads%
\begin{equation*}
\frac{\partial (\mathbf{w}\cdot \nabla \phi )}{\partial t}+\nabla \cdot %
\left[ \nabla \phi \times \left( \mathbf{v}\times \mathbf{w}+\nu \nabla
\times \mathbf{w}\right) -\frac{\partial \phi }{\partial t}\mathbf{w}\right]
=0
\end{equation*}%
which can be put into the form%
\begin{equation*}
\frac{d}{dt}\left( \mathbf{w}\cdot \nabla \alpha \right) -\mathbf{w}\cdot
\nabla \frac{d\phi }{dt}-\nu \nabla \phi \cdot \nabla ^{2}\mathbf{w}=0
\end{equation*}%
with the convective time derivative $d/dt$.
\end{proof}
\end{proposition}

\textbf{10. Symplectic Dilation: }A particularly interesting solution of Eq.(%
\ref{flat}) is obtained when we let the one-form $\beta $ be the canonical
one-form $\theta $. The vector field $J_{\nu }$ satisfying the equation 
\begin{equation}
i(J_{\nu })(\Omega _{\nu })=\theta  \label{teta}
\end{equation}%
can be uniquely determined to be 
\begin{equation}
J_{\nu }={\frac{1}{\mathbf{w}\cdot \nabla \phi -\nu \mathcal{H}_{w}}}[%
\mathcal{H}(\partial _{t}+\mathbf{v}\cdot \nabla )-(\psi +v^{2})\mathbf{w}%
\cdot \nabla +\mathbf{v}\times (\nabla \phi -\nu \nabla \times \mathbf{w)}%
\cdot \nabla ]\;.  \label{dil}
\end{equation}%
It follows from Eq.(\ref{teta}) and $d\Omega _{\nu }=0$ that $J_{\nu }$
fulfills the condition 
\begin{equation}
\mathcal{L}_{J_{\nu }}(\Omega _{\nu })=di(J_{\nu })(\Omega _{\nu })=d\theta
=\Omega _{\nu }  \label{ome}
\end{equation}%
of being an infinitesimal symplectic dilation for $\Omega _{\nu }$ \cite%
{wei76}. As a consequence of Eq.(\ref{ome}) and the derivation property of
the Lie derivative we see that $J_{\nu }$ expands the Liouville volume in
Eq.(\ref{vol}). That means, $J_{\nu }$ is not divergence free with respect
to the symplectic volume. $J_{\nu }$ is also said to be the Liouville vector
field of $\Omega _{\nu }$ \cite{LM}. The symplectic divergence of the
dilation $J_{\nu }$ may be computed from 
\begin{equation}
(div_{\Omega _{\nu }}J_{\nu })({\frac{1}{2}}\Omega _{\nu }\wedge \Omega
_{\nu })=\mathcal{L}_{J_{\nu }}({\frac{1}{2}}\Omega _{\nu }\wedge \Omega
_{\nu })=\Omega _{\nu }\wedge \Omega _{\nu }
\end{equation}%
where we used Eq.(\ref{ome}). The second equality is the same as the
identity in Eq.(\ref{tom}) resulting in the helicity evolution. Thus, we have

\begin{proposition}
The evolution equation in Eq.(\ref{orhe}) for helicity density can be
expressed by the equation 
\begin{equation}
div_{\Omega _{\nu }}J_{\nu }-2\equiv 0
\end{equation}%
involving the symplectic-divergence of the symplectic dilation in Eq.(\ref%
{dil}).
\end{proposition}

With this interpretation we intend to call $J_{\nu }$ to be the current
associated with the helicity. The dynamical content of the helicity current
can be revealed from a comparison of the symplectic structures obtained from
the Navier-Stokes and the Euler equations. The canonical one-forms are the
same. So, the dynamics is encoded into the symplectic two-forms. They define
the current vectors by Eq.(\ref{teta}) for the same canonical one-form. With
this definition, the dynamical properties of the fluid, such as viscosity,
become implicit in the helicity current. Thus, we can conclude that the pairs%
\begin{equation*}
(\theta ,J_{e}),\text{ \ \ \ \ \ \ \ \ \ \ \ \ }(\theta ,J_{\nu })
\end{equation*}%
are geometric representatives of the dynamics of ideal and viscous fluid
motions on the space of trajectories.

\textbf{11. Hamiltonian Automorphisms: }The helicity current is not a
Hamiltonian vector field. However, it takes a Hamiltonian vector field into
a Hamiltonian vector field by its action via Lie derivative. To see this, we
compute 
\begin{eqnarray}
i([J_{\nu },X_{f}])(\Omega _{\nu }) &\equiv &\mathcal{L}_{J_{\nu
}}(i(X_{f})(\Omega _{\nu }))-i(X_{f})(\mathcal{L}_{J_{\nu }}(\Omega _{\nu }))
\label{jx} \\
&=&d(J_{\nu }(f)-f)  \label{jfx}
\end{eqnarray}%
where Eq.(\ref{jx}) is an identity \cite{AMR},\cite{MR} and we used Eq.(\ref%
{ome}). Thus, $[J_{\nu },X_{f}]$ is Hamiltonian with the function $J_{\nu
}(f)-f$. Replacing $X_{f}$ with $[J_{\nu },X_{f}]$ in Eq.(\ref{jx}) and
using Eq.(\ref{jfx}) we get 
\begin{eqnarray}
i([J_{\nu },[J_{\nu },X_{f}]])(\Omega _{\nu }) &\equiv &\mathcal{L}_{J_{\nu
}}(i([J_{\nu },X_{f}])(\Omega _{\nu }))-i([J_{\nu },X_{f}])(\mathcal{L}%
_{J_{\nu }}(\Omega _{\nu })) \\
&=&d(J_{\nu }(J_{\nu }(f))-2J_{\nu }(f)+f) \\
&=&d(J_{\nu }-1)^{2}(f)
\end{eqnarray}%
which is also Hamiltonian. Thus, by repeated applications of the Lie
derivative with respect to the symplectic dilation $J_{\nu }$ one can
generate an infinite hierarchy of Hamiltonian vector fields.

\begin{proposition}
Let $X_{f}$ be a Hamiltonian vector field $i(X_{f})(\Omega _{\nu })\equiv df$%
. Then, for each $k=0,1,2,...$ the vector fields $(\mathcal{L}_{J_{\nu
}})^{k}(X_{f})$ are Hamiltonian with respect to $\Omega _{\nu }$ and for the
Hamiltonian functions $(J_{\nu }-1)^{k}(f)$.
\end{proposition}

This infinite hierarchy of Hamiltonian vector fields are anchored to $X_{f}$%
. Since%
\begin{equation*}
\mathcal{L}_{[J_{\nu },X_{f}]}(\Omega _{\nu })\equiv \mathcal{L}_{J_{\nu }}%
\mathcal{L}_{X_{f}}(\Omega _{\nu })-\mathcal{L}_{X_{f}}\mathcal{L}_{J_{\nu
}}(\Omega _{\nu })=0
\end{equation*}%
they are Hamiltonian automorphisms of $\Omega _{\nu }$ on $\mathcal{%
\mathbb{R}
\times }M$.

\section{Lagrangian Descriptions}

\textbf{12. Hamiltonian Velocity Fields: }The formal symplectic structure
obtained from the Eulerian equations immediately implies that the normalized
vorticity field is Hamiltonian on $\mathcal{%
\mathbb{R}
\times }M$ with the Hamiltonian function $t$. Our real interest is the
geometry of the velocity field $\mathbf{v}$ or its suspension $\partial _{t}+%
\mathbf{v}\cdot \nabla $. With respect to the two-form $\Omega _{\nu }$ the
suspended velocity field is, in general, not even locally Hamiltonian.
(Locally Hamiltonian vector fields are obtained from Eq.(\ref{flat}) for
closed non-exact one-forms $\beta )$. To see this, first observe that, the
symplectic two-form is invariant under the flows of locally Hamiltonian
vector fields because $\mathcal{L}_{X}(\Omega _{\nu })=di(X)(\Omega _{\nu
})=d\beta \equiv 0$ where we used the identity $\mathcal{L}_{X}=i(X)\circ
d+d\circ i(X)$ for the Lie derivative, $d\Omega _{\nu }=0$ and $\beta $ is
closed.

\begin{proposition}
The suspended velocity field $\partial _{t}+v$ is Hamiltonian 
\begin{equation}
i(\partial _{t}+v)(\Omega _{\nu })=-d\alpha  \label{hamsus}
\end{equation}%
with the Hamiltonian function being the negative of the Bernoulli's function
whenever $\mathbf{w}$ is a frozen in field.

\begin{proof}
To obtain the conditions under which $\partial _{t}+v$ is Hamiltonian we
compute, from Eq.(\ref{nsym}), 
\begin{eqnarray}
i(\partial _{t}+v)(\Omega _{\nu }) &=&(\mathbf{v}\times \mathbf{w}+\nabla
\phi +\nu \nabla ^{2}\mathbf{v})\cdot d\mathbf{x} \\
&&\mathbf{+\mathbf{w}\times \mathbf{v}\cdot }d\mathbf{\mathbf{x}-v\cdot (}%
\nabla \phi +\nu \nabla ^{2}\mathbf{v)}dt \\
&=&(\nabla \phi +\nu \nabla ^{2}\mathbf{v})\cdot (d\mathbf{x}-\mathbf{v}dt)
\\
&=&\mathbf{-(v\cdot }\nabla (\phi +\alpha )+\frac{\partial v^{2}/2}{\partial
t}\mathbf{)}dt+(\nabla \phi +\nu \nabla ^{2}\mathbf{v})\cdot d\mathbf{x} \\
&=&-d\alpha +(\frac{\partial p}{\partial t}+\mathbf{v}\cdot \nabla \psi
)dt+(\nu \nabla ^{2}\mathbf{v}-\nabla \psi )\cdot d\mathbf{x}
\label{suspend}
\end{eqnarray}%
where in the third line we used the Navier-Stokes equations to replace $\nu
\nabla ^{2}\mathbf{v}$ in coefficient of $dt$ and $-\psi =\phi +\alpha $. If
we require the right hand side to be an exact one-form, the second and the
third terms must be time derivative and gradient of some function,
respectively. But we have already introduced the arbitrary function $\phi $,
which implies the arbitrariness of $\psi $. So, we set 
\begin{equation}
\frac{\partial p}{\partial t}+\mathbf{v}\cdot \nabla \psi =0\text{ , \ \ \ \
\ \ \ \ \ \ }\nu \nabla ^{2}\mathbf{v}-\nabla \psi =0  \label{psi}
\end{equation}%
as the condition for the right hand side of Eq.(\ref{suspend}) to be exact.
The integrability condition $\nabla \times \nabla \psi =0$ implies $\nu
\nabla ^{2}\mathbf{w=}0$. This holds in cases of either the ideal flow $\nu 
\mathbf{=}0$ or, $\nabla ^{2}\mathbf{w=}\nabla \times \nabla \times \mathbf{w%
}=0$ for the viscous case. In either cases, $\mathbf{w}$ becomes a frozen in
field by the vorticity form of Euler and Navier-Stokes equations,
respectively. Elimination of the function $\psi $ in Eq.(\ref{psi}) leads to
Eq.(\ref{pressure}) which quarantees the conservation of Bernoulli's
function $\alpha $ under the time-dependent flow of velocity field.
\end{proof}
\end{proposition}

\textbf{13. Differential Invariants:} The symplectic structure provides the
velocity field with differential invariants which in turn enable us to
obtain Eulerian conservation laws. Recall that a differential $p-$form $\xi $
is said to be a relative invariant for a vector field $X$ if there exist a $%
(p-1)-$form $\zeta $ such that 
\begin{equation}
\mathcal{L}_{X}(\xi )\equiv i(X)d\xi +di(X)\xi =d\zeta  \label{inv}
\end{equation}%
where $\mathcal{L}_{X}(\cdot )$ is the Lie derivative. If $\zeta =0$, $\xi $
is said to be an absolute invariant \cite{LM},\cite{AMR}.

\begin{proposition}
Let $\phi =-\alpha $, or equivalently $\psi =0$. Then, $\theta $ and $\theta
\wedge \Omega _{\nu }$ become absolute invariants of $\partial _{t}+v$ on
level surfaces of the function $v^{2}/2-p$. In this case, the helicity
density is a conserved function of the Lagrangian motion.

\begin{proof}
From Eq.(\ref{suspend}) and assuming that Eq.(\ref{psi}) holds we have%
\begin{eqnarray}
\mathcal{L}_{\partial _{t}+v}(\theta ) &=&i(\partial _{t}+v)(\Omega _{\nu
})+d(v^{2}-\phi -\alpha ) \\
&=&d(\frac{v^{2}}{2}-p-\phi -\alpha ),
\end{eqnarray}%
that is, $\theta $ is a relative invariant for incompressible fluid
satisfying the conditions for the Bernoulli's function to be conserved. In
this case, $\mathcal{L}_{\partial _{t}+v}(\Omega _{e})=d\mathcal{L}%
_{\partial _{t}+v}(\theta )=0$ and the derivation property of the Lie
derivative implies the relative invariance 
\begin{eqnarray}
\mathcal{L}_{\partial _{t}+v}(\theta \wedge \Omega _{\nu }) &=&\mathcal{L}%
_{\partial _{t}+v}(\theta )\wedge \Omega _{\nu }+\theta \wedge \mathcal{L}%
_{\partial _{t}+v}(\Omega _{\nu })  \label{inv3} \\
&=&d((\frac{v^{2}}{2}-p-\phi -\alpha )\Omega _{\nu })
\end{eqnarray}%
of the three-form $\theta \wedge \Omega _{\nu }$. Recall that the function $%
\phi $ was arbitrarily introduced into $\Omega _{\nu }$ and its dynamical
significance is yet to be specified. To this end, it will be convenient to
let $\phi =-\alpha $ in accordance with the fact that the helicity evolution
is independent of this function. Then, both $\theta $ and $\theta \wedge
\Omega _{\nu }$ becomes an absolute invariant on surfaces $v^{2}/2-p=$%
constant. The condition for relative invariance makes the helicity density
an Eulerian conserved quantity (c.f. Eq.(\ref{orhe})), while on surfaces $%
v^{2}/2-p=$constant the helicity density becomes a conserved function of the
Lagrangian motion.
\end{proof}
\end{proposition}

For ideal fluids, the potential vorticity $\mathbf{w}\cdot \nabla \alpha $
is also conserved under the flow of velocity field. From the right hand side
of Eq.(\ref{vol}) we conclude that the absolute invariance of $\Omega
_{e}\wedge \Omega _{e}$ is a statement for the conservation of the Liouville
density 
\begin{equation}
{\frac{\partial }{\partial t}}(\mathbf{w}\cdot \nabla \alpha )+\mathbf{v}%
\cdot \nabla (\mathbf{w}\cdot \nabla \alpha )=0  \label{conl}
\end{equation}%
along the trajectories of the velocity field.

\section{Symplectic Dilation and Contact Structures}

We shall describe two examples of contact hypersurfaces in $\mathcal{%
\mathbb{R}
\times }M$. Then we shall consider their relation with the symplectic
structure $\Omega _{\nu }$ and discuss some physical significance. In this
section, we will assume $\mathcal{H}_{w}=0$ and $\phi =-\alpha $.

A contact structure on a three dimensional manifold is a field of
non-integrable, two-dimensional hyperplanes in its tangent spaces. Locally,
this may be described as the kernel of a one-form $\sigma $ satisfying $%
\sigma \wedge d\sigma \neq 0$ everywhere. The contact form $\sigma $
determines a unique vector field $E$ by the conditions 
\begin{equation}
i(E)(\sigma )=1\;,\;\;\;i(E)(d\sigma )=0  \label{reeb}
\end{equation}%
which is called the Reeb vector field \cite{ARN},\cite{DUSA}. According to
the definition of Ref. \cite{wei79} a hypersurface in a symplectic manifold
admits a contact one-form if and only if there exists a symplectic dilation
which is defined on its neighborhood and is transversal to the hypersurface.
It is also remarked that such hypersurfaces arise as level sets of
Hamiltonian functions of Hamiltonian vector fields and that the periodic
solutions of Hamiltonian vector fields exist on contact hypersurfaces \cite%
{wei76},\cite{DUSA},\cite{rab79}.

\textbf{14. Spatial Hypersurfaces:} First, we consider spatial hypersurfaces 
\begin{equation*}
M_{c}=\{(t,x)\in \mathcal{%
\mathbb{R}
\times }M\;|\;t=c=constant\}
\end{equation*}%
as level sets of the function $t$. Recall that $t$ is the Hamiltonian
function for the normalized vorticity field $X_{t}=(\mathbf{w}\cdot \alpha
)^{-1}\mathbf{w}\cdot \nabla $. Then, the transversality condition 
\begin{equation}
i(J_{\nu })(i(X_{t})(\Omega _{\nu }))=i(J_{\nu })(dt)={\frac{2\mathcal{H}}{%
\mathbf{w}\cdot \nabla \alpha }}\neq 0
\end{equation}%
for $M_{c}$ holds for non-vanishing values of the helicity density. In other
words, for $\mathcal{H}\neq 0$ the helicity current is not contained in the
tangent spaces to $M_{c}$. This is obvious from the explicit expression in
Eq.(\ref{dil}) for $J$ in which $\mathcal{H}$ occurs as the component in the
time direction. In this case, the contact form on $M_{c}$ is obtained as
follows. Let $i:M_{c}\rightarrow \mathcal{%
\mathbb{R}
\times }M:(t=c,x)\mapsto (t,x)$ be the inclusion of time slices $M_{c}$ into
space-time. A function on $\mathcal{%
\mathbb{R}
\times }M$ gives a function on $M_{c}$ when composed by $i$. This operation
can then be extended to differential forms. If $\gamma =f_{a}(x)dx^{a}$ is a
one-form on $\mathcal{%
\mathbb{R}
\times }M$ its pull-back $i^{\ast }\gamma $ to $M_{c}$ by the inclusion map $%
i$ is defined to be 
\begin{equation}
i^{\ast }\gamma =i^{\ast }(f_{a}(x)dx^{a})=i^{\ast }(f_{a}(x))i^{\ast
}(dx^{a})=(f_{a}\circ i)(x)di(x^{a})
\end{equation}%
where we used the commutativity of the operators $d$ and $i^{\ast }$ \cite%
{AMR}.

\begin{proposition}
On the spatial hypersurfaces $M_{c}$, the one-form 
\begin{equation}
\sigma \equiv i^{\ast }\theta \;=\;\mathbf{v}(t=c,\mathbf{x})\cdot d\mathbf{%
x=}i(J_{\nu })(\Omega _{\nu })|_{M_{c}}  \label{cont}
\end{equation}%
defines a contact structure.
\end{proposition}

In particular, we compute 
\begin{equation*}
d\sigma =di^{\ast }\theta \;=\;i^{\ast }d\theta =i^{\ast }\Omega _{\nu }=%
\mathbf{w}(t=c,\mathbf{x})\cdot d\mathbf{x}\wedge d\mathbf{x}
\end{equation*}%
and it follows that $\sigma $ is a contact form on $M_{c}$ 
\begin{equation}
\sigma \wedge d\sigma =i^{\ast }\theta \wedge i^{\ast }\Omega _{\nu
}=i^{\ast }(\theta \wedge \Omega _{\nu })=2\mathcal{H}\,d\mathbf{x}\cdot d%
\mathbf{x}\wedge d\mathbf{x}\neq 0
\end{equation}%
provided the helicity density is non-zero. Note that as a result of the
transversality condition we have $i(J_{\nu })(\sigma )=0$. In fact, $%
i(J_{\nu })(\theta )=0$ follows from the very definition of the symplectic
dilation $J_{\nu }$. Projections on $M_{c}$ of equilibrium solutions for the
Lagrangian motion corresponds to Legendrian submanifolds of the contact
structure.

We shall now show that the investigation of the normalized vorticity field $%
X_{t}$ as a Hamiltonian system on $\mathcal{%
\mathbb{R}
\times }M$ is equivalent to the study of the Reeb vector field on the level
surfaces $M_{c}$ of the Hamiltonian function $t$. Since $t\circ i$ is the
constant function on time slices $M_{c}$ we have 
\begin{equation}
i^{\ast }(i(X_{t})(\Omega _{\nu }))=i^{\ast }(dt)=di^{\ast }t=d(t\circ i)=0
\label{hom}
\end{equation}%
as the pull-back of the Hamilton's equations for $X_{t}$ to the spatial
hypersurfaces. On the other hand, for any vector field $X$ on $\mathcal{%
\mathbb{R}
\times }M$ we have the identity 
\begin{equation}
i^{\ast }(i(X)(\Omega _{\nu }))=i(i_{\ast }(X))(i^{\ast }\Omega _{\nu
})=i(i_{\ast }(X))(d\sigma )
\end{equation}%
where $i_{\ast }X$ denotes the push-forward of $X$ to $M_{c}$. That means, $%
i_{\ast }X$ is the pull-back of $X$ by $i^{-1}$ and hence is a vector field
on $M_{c}$ \cite{AMR}. Since the one dimensional kernel of $d\sigma $ (in
the tangent spaces of $M_{c}$) is spanned by the Reeb vector field, Eq.(\ref%
{hom}) for $X_{t}$ is possible only if the push-forward $i_{\ast }X_{t}$ is
proportional to the Reeb vector field of the contact structure on $M_{c}$.
In fact, it is easy to check that the vector field 
\begin{equation}
\mathbf{E}(c,\mathbf{x})={\frac{1}{\mathcal{H}}}\mathbf{w}(t=c,\mathbf{x})=({%
\frac{\mathbf{w}\cdot \nabla \alpha }{\mathcal{H}}}\mathbf{X}_{t})(t=c,%
\mathbf{x})
\end{equation}%
with $\mathcal{H}$ being evaluated at $t=c$ satisfies the criteria in Eq.(%
\ref{reeb}) for the contact one form in Eq.(\ref{cont}).

\textbf{15. Comparison with Steady Flow: }To this end, we want to remark
that the contact structure on spatial hypersurfaces $M_{c}$ must be
distinguished from similar geometric constructions on $M$ obtained from the
Euler equation 
\begin{equation}
\mathbf{v}(\mathbf{x})\times \mathbf{w}(\mathbf{x})=\nabla \alpha (\mathbf{x}%
)  \label{seuler}
\end{equation}%
for the steady flow of incompressible fluid. In this latter case there is
also a contact structure on $M$ provided the (time-independent) helicity
density is non-zero. However, the difference between the two cases is not
merely the time-dependence of fields. They imply qualitatively different
descriptions of the flows. For example, from Eq.(\ref{seuler}) of the steady
flow we obtain 
\begin{equation}
\mathbf{v}(\mathbf{x})\cdot \nabla \alpha (\mathbf{x})=0\;,\;\;\;\mathbf{w}(%
\mathbf{x})\cdot \nabla \alpha (\mathbf{x})=0\;,\;\;\;[\mathbf{v}(\mathbf{x}%
),\mathbf{w}(\mathbf{x})]=0
\end{equation}%
which means that the fields $\mathbf{v}$ and $\mathbf{w}$ span the tangent
spaces of the (two dimensional) Bernoulli surfaces $\alpha (\mathbf{x}%
)=constant$ and their flow lines commute on these surfaces \cite{ARN}. On
the other hand, the pull-back of the unsteady Euler equation to the
hypersurfaces $M_{c}$ gives 
\begin{equation}
\mathbf{v}_{,t}(t=c,\mathbf{x})-\mathbf{v}(t=c,\mathbf{x})\times \mathbf{w}%
(t=c,\mathbf{x})=\nabla \alpha (t=c,\mathbf{x})\;.  \label{mce}
\end{equation}%
The qualitative analysis of this equation implies quite different and
complicated results for the surfaces $\alpha (t=c,\mathbf{x})=constant$ as
well as the expression for the push-forward of the suspended velocity field
to the contact hypersurfaces $M_{c}$. The behaviour of the flows lines on $%
M_{c}$ are also different. For example, if we take the curl of Eq.(\ref{mce}%
) 
\begin{equation}
\mathbf{w}_{,t}(t=c,\mathbf{x})+[\mathbf{v}(t=c,\mathbf{x}),\mathbf{w}(t=c,%
\mathbf{x})]=0
\end{equation}%
we see that the flows of the velocity and the vorticity fields do not
necessarily commute on $M_{c}$.

\textbf{16.} \textbf{Bernoulli Hypersurfaces: }Let $B\subset \mathcal{%
\mathbb{R}
\times }M$ be the level sets of the Bernoulli function $\alpha $ which is
the Hamiltonian function for the suspended velocity field of the Euler flow.
Let $i:B\rightarrow \mathcal{%
\mathbb{R}
\times }M$ be the inclusion. The transversality condition reads 
\begin{equation}
i(X_{\alpha })(\theta )=i(J)(d\alpha )=-v^{2}\neq 0
\end{equation}%
where we used the fact that $\alpha $ is conserved under the Lagrangian
motion. The transversality of the helicity current to the Bernoulli surfaces
is the same as the non-vanishing of the kinetic energy of the fluid. On $B$,
the Euler equation becomes 
\begin{equation}
{\frac{\partial \mathbf{v}}{\partial t}}-\mathbf{v}\times \mathbf{w}%
=0\;,\;\;\;\alpha =constant\;.  \label{bere}
\end{equation}%
Obviously, the pull-back of the symplectic two-form 
\begin{equation}
i^{\ast }\Omega _{e}=\mathbf{w}\cdot (d\mathbf{x}\wedge d\mathbf{x})-\mathbf{%
v}\times \mathbf{w}\cdot d\mathbf{x}\wedge dt  \label{bsym}
\end{equation}%
to the Bernoulli surfaces is degenerate. Its one-dimensional kernel in the
tangent spaces of $B$ is the span of the vorticity field. Since $\partial
_{t}+v$ is also tangent to $B$ the two-dimensional tangent hypersurfaces (in
the three dimensional tangent spaces of $B$) on which $i^{\ast }\Omega _{e}$
is non-degenerate can be defined to be the complement of $span\{\partial
_{t}+v,\mathbf{w}\cdot \nabla \}$ in the tangent spaces of $\mathcal{%
\mathbb{R}
\times }M$. $i^{\ast }\Omega _{e}$ is also exact on $B$ 
\begin{equation}
i^{\ast }\Omega _{e}=d(\mathbf{v}\cdot d\mathbf{x})\;\;mod\;Eq.(\ref{bere})
\end{equation}%
by the pulled-back of Euler equation. So, the contact structure on the
Bernoulli surfaces is defined by the time-dependent one-form 
\begin{equation}
\sigma _{t}=\mathbf{v}(t,\mathbf{x})\cdot d\mathbf{x}
\end{equation}%
whose derivative is the two-form in Eq.(\ref{bsym}). The non-integrability
condition 
\begin{equation}
\sigma _{t}\wedge d\sigma _{t}=2\mathcal{H}\;d\mathbf{x}\cdot d\mathbf{x}%
\wedge d\mathbf{x}+(v^{2}\mathbf{w}-2\mathcal{H}\mathbf{v})\cdot d\mathbf{x}%
\wedge d\mathbf{x}\wedge dt
\end{equation}%
of the tangent hyperplanes defined as above requires either the helicity
density or the kinetic energy to be non-zero.

\begin{proposition}
Let $B=\{(t,\mathbf{x})\in \mathcal{%
\mathbb{R}
\times }M:\alpha =p(\mathbf{x})+v^{2}(t,\mathbf{x})/2=$constant$\}$ be level
sets of Bernoulli surfaces. Following are equivalent:

i) the kinetic energy is non-zero.

ii) $J_{\nu }$ is transversal to $B.$

iii) $\sigma _{t}$ defines a contact structure on $B$.
\end{proposition}

Recall that the non-vanishing of the kinetic energy is also required by the
transversality condition. For the contact one-form $\sigma _{t}$ on the
Bernoulli surfaces we can find the Reeb vector field up to an arbitrary
function 
\begin{equation}
E=m(t,x)(\partial _{t}+v)+n(t,x)\mathbf{w}\cdot \nabla \;,\;\;\;mv^{2}+2n%
\mathcal{H}=1\;.
\end{equation}%
This arbitrariness is a manifestation of the fact that contrary to the case
of spatial hypersurfaces the inclusion of the Bernoulli hypersurfaces into $%
\mathcal{%
\mathbb{R}
\times }M$ is defined only implicitly.

\section{Discussion and Conclusions}

\textbf{17.} \textbf{Symplectisation: }The geometric fluid dynamics provides
an unusual but nevertheless a natural example of symplectisation. Recall
that we obtain the contact one-form on time slices by pulling the canonical
one-form $\theta $ back to $M_{c}$ by the inclusion. Conversely, the
symplectisation of the contact structure on time slices follows from the
inclusion map 
\begin{equation}
i:M_{c}\rightarrow \mathcal{%
\mathbb{R}
\times }M:(t=c,\mathbf{x})\mapsto (t,\mathbf{x})\;.
\end{equation}%
In this case, the time variable $t$ is introduced naturally by the action of
the invariant differential operators.

From a physical point of view the symplectisation of the time slices $M_{c}$
corresponds to the construction of trajectories of the velocity field from
streamlines. These are solutions of the non-autonomous and autonomous
equations 
\begin{equation}
{\frac{d\mathbf{x}(t)}{dt}}=\mathbf{v}(t,\mathbf{x})\;,\;\;\;{\frac{d\mathbf{%
x}(\tau )}{d\tau }}=\mathbf{v}(t=c,\mathbf{x}(\tau ))\;,  \label{difeq}
\end{equation}%
respectively. The solutions to the first equation on $\mathcal{%
\mathbb{R}
\times }M$ can be constructed by solving the autonomous system on $M$ at
each time $t$ and then joining them by the inclusion 
\begin{equation}
i:(streamlines)\mapsto (trajectories)\;.  \label{st}
\end{equation}%
The symplectisation to $\mathcal{%
\mathbb{R}
\times }M$ of the contact structures on $M_{c}$ means that the inclusion in
Eq.(\ref{st}) for solutions of the differential equations (\ref{difeq})
extends to the inclusion of geometric structures 
\begin{equation*}
\begin{array}{cccc}
i\;: & \left( 
\begin{array}{c}
contact\;structure\;on \\ 
the\;space\;of\;streamlines%
\end{array}%
\right) & \longrightarrow & \left( 
\begin{array}{c}
symplectic\;structure\;on \\ 
the\;space\;of\;trajectories%
\end{array}%
\right)%
\end{array}%
\end{equation*}%
on $M_{c}$ into those on $\mathcal{%
\mathbb{R}
\times }M$.

The symplectisation of the contact structure on the Bernoulli surfaces may
be given a similar interpretation in the language of the solutions of
differential equations. In this case, it will be appropriate to consider the
solutions of the Euler equation. The inclusion 
\begin{equation*}
\begin{array}{cccc}
i\;: & \left( 
\begin{array}{c}
Bernoulli \\ 
surfaces%
\end{array}%
\right) & \longrightarrow & \left( 
\begin{array}{c}
space-time%
\end{array}%
\right)%
\end{array}%
\end{equation*}%
implies the construction 
\begin{equation*}
\begin{array}{cccc}
i\;: & \left( 
\begin{array}{c}
solution\;of \\ 
\mathbf{v}_{,t}-\mathbf{v}\times \mathbf{w}=0 \\ 
on\;\alpha =constant%
\end{array}%
\right) & \longrightarrow & \left( 
\begin{array}{c}
solution\;of \\ 
the\;Euler\;equation%
\end{array}%
\right)%
\end{array}%
\end{equation*}%
of the solutions of the Euler equation from the solutions of a homogeneous
equation on each Bernoulli surface. This may be interpreted as a nonlinear
analog of the result in the theory of linear differential equations that the
general solution of an inhomogeneous equation is the sum of the solution of
its homogeneous part and the particular solution.

\bigskip

\end{document}